\documentclass{article}
\usepackage{times}
\usepackage{amsfonts}
\usepackage{graphicx}
\begin{document}
\noindent
{\Large  HYPERBOLIC SPACE IN THE NEWTONIAN LIMIT:\\ THE COSMOLOGICAL CONSTANT}
\noindent

\vskip.5cm
\noindent
{\bf J.C. Castro-Palacio}$^{1,a}$, {\bf P. Fern\'andez de C\'ordoba}$^{2,b}$,  {\bf R. Gallego Torrom\'e}$^{2,c}$\\ and {\bf  J.M. Isidro}$^{2,d}$\\
$^{1}$Centro de Tecnolog\'{\i}as F\'{\i}sicas: Ac\'ustica, Materiales y Astrof\'{\i}sica,\\Universitat Polit\`ecnica de Val\`encia, Valencia 46022, Spain\\
$^{2}$Instituto Universitario de Matem\'atica Pura y Aplicada,\\ Universitat Polit\`ecnica de Val\`encia, Valencia 46022, Spain\\
$^{a}${\tt juancas@upvnet.upv.es}, $^{b}${\tt pfernandez@mat.upv.es},\\
$^{c}${\tt rigato39@gmail.com}, $^{d}${\tt joissan@mat.upv.es}\\

\vskip.5cm
\noindent
\today
\vskip.5cm
\noindent
{\bf Abstract}  The cosmological constant and the Boltzmann entropy of a Newtonian Universe filled with a perfect fluid are computed, under the assumption that spatial sections are copies of 3--dimensional hyperbolic space.

\section{Introduction}\label{einfuehrung}

{}For as long as the nature of the cosmological constant $\Lambda$ remains a mystery \cite{CARROLL, MARTIN, PEEBLES,  PERLMUTTER, RIESS, WEINBERG1} one can turn things around and try to explain related issues first, in a search for possible hints to understand the underlying physics  fully at a later step. In this spirit, the implications of a positive cosmological constant have been analysed in ref. \cite{ASHTEKAR}. 

Very reasonably, a complete understanding of the nature of the cosmological constant requires a consistent framework embracing together a quantum mechanical description and the gravitational interaction. From this point of view, new paradigms have been recently proposed where quantum mechanics is emergent and gravity appears as a classical, emergent interaction \cite{RICARDO2014, RICARDO2015, RICARDO2020, SINGH2020}. Gravity is classical in such frameworks, as in refs. \cite{JACOBSON, VERLINDE1}. All these proposals are of such a radical nature that one expects, once understood, they will offer powerful insights into deep current open questions in cosmology. In this context of classical gravity paradigms, the problem about the nature and properties of the cosmological constant arises as a natural question to be considered. In particular, in this paper we are concerned with the consequences that the entropic point of view of gravity has on the cosmological constant.

Entropic notions play a central role in novel approaches to key problems in theoretical physics \cite{FINSTER1, FINSTER2}.  In refs. \cite{PADDY1, PADDY2}, a nonzero value of the cosmological constant has been argued to render finite the amount of cosmic information accessible to an eternal observer.  An explicit proposal concerning the possible nature of the atoms of spacetime has been put forward in ref. \cite{SINGH}.  The entropic approach to gravity pioneered in refs. \cite{VERLINDE1, VERLINDE2}, as well as the holographic bound, have been exploited in refs. \cite{CABRERA, NOSALTRES, NOSALTRESTRIS, JMI}.  Altogether we see that a lot of useful information can be extracted via this indirect approach, where one tries to probe the fabric of the cosmos by means of simpler, or possibly better known, physical systems. 

One such simplification is provided by the Newtonian limit \cite{BARROW1, BENISTY1, GIBBONS, BENISTY2, VIGNERON1, VIGNERON2, VIGNERON3}, as opposed to the fully relativisitic regime. Further advantage is provided by {\it the duality between the Newtonian cosmological fluid on the one hand, and the probability fluid of nonrelativistic quantum mechanics, on the other}\/. This duality has been first exhibited in our papers \cite{CABRERA, NOSALTRESTRIS}. It provides an equivalent, but more manageable description of the cosmological fluid, because it uses the well--known operator formalism of nonrelativistic quantum theory even if the system under discussion is classical instead of quantum. Although no conceptual simplification is afforded by this duality, the formalism of quantum operators renders the approach computationally straightforward. At the end of the day, one simply ignores the quantum operators used at intermediate steps of the calculations and replaces them with their expectation values in suitably selected quantum states. If the latter enjoy semiclassical properties, then the use of quantum theory for obtaining a classical result is justified {\it a posteriori}\/.

We have applied the above approach to a computation of the cosmological constant of flat space in ref. \cite{NOSALTRES}, and to the Einstein Universe in ref. \cite{JMI}.  These toy Universes are respectively described by the manifolds  $\mathbb{R}\times\mathbb{R}^3$ and $\mathbb{R}\times\mathbb{S}^3$. Here the real line stands for the time variable, while $\mathbb{S}^3$ is a 3--dimensional sphere endowed with the usual round metric.   

In this paper we extend the analysis of refs. \cite{NOSALTRES, JMI} to the case when the spatial manifold is $\mathbb{H}^3$, 3--dimensional hyperbolic space. Our Newtonian Universe is thus $\mathbb{R}\times\mathbb{H}^3$, with the real line standing for the time axis. In this way we exhaust all possible cases for a homogeneous, isotropic spacetime manifold as a realistic representation of our actual Universe (within the Newtonian limit, and modulo quotients by discrete groups, which will not be considered here). As before, our aim is to compute the numerical values of the cosmological constant and of the Boltzmann entropy of this Universe. Quantum calculations in hyperbolic space, including corrections to the cosmological constant, were reported long ago in refs. \cite{BYTSENKO1, BYTSENKO2}.

As a solution to the Einstein field equations, hyperbolic space carries a negative value of the cosmological constant \cite{WEINBERG2}. However, our result for $\Lambda$ is determined only up to a dimensionless real factor $C_\Lambda$.  Although, on general grounds, $C_\Lambda$ is expected to be of order unity, the choice of sign remains arbitrary. Here we will show that a choice of $C_\Lambda>0$ exists such that it reproduces the experimentally determined value of $\Lambda>0$. In so doing one is {\it not}\/ contradicting the fact that hyperbolic space carries $\Lambda<0$, since $C_{\Lambda}$ remains arbitrary. Actually the real Universe is flat but has $\Lambda>0$. The same ambiguity in the sign of $\Lambda$, due to an undetermined dimensionless factor, already arose in our analyses of flat space \cite{NOSALTRES} and of spherical space \cite{JMI}.   

In our use of special functions we follow the conventions of ref. \cite{GRADSHTEYN}. In the notation $X_{\rm reg}$, the subscript ${}_{\rm reg}$ indicates that the divergent quantity $X$ has been regularised as indicated in the text.

\section{The Laplacian in hyperbolic space}\label{hiperbolico}

\subsection{Laplacian eigenfunctions}

Consider flat space $\mathbb{R}^4$ endowed with Cartesian coordinates $x^j$, $j=1,2,3,4$, and the Minkowski metric $(+,+,+,-)$.
The equation 
\begin{equation}
(x^1)^2+(x^2)^2+(x^3)^2-(x^4)^2=-R_0^2
\label{uno}
\end{equation}
defines a 3--dimensional hyperboloid with radius $R_0>0$; either one of its two sheets is a copy of 3--dimensional hyperbolic space $\mathbb{H}^3$. 
A convenient parametrisation of $\mathbb{H}^3$ is given by $r\in[0,\infty)$, $\theta\in[0,\pi/2]$ and $\varphi\in[0,2\pi)$, where
\begin{eqnarray}
x^1&=&R_0\sinh r\sin\theta\cos\varphi\nonumber\\
x^2&=&R_0\sinh r\sin\theta\sin\varphi\nonumber\\
x^3&=&R_0\sinh r\cos\theta  \nonumber\\
x^4&=&R_0\cosh r.
\label{dos}
\end{eqnarray}
In these coordinates the induced metric on $\mathbb{H}^3$ reads
\begin{equation}
{\rm d}s^2=R_0^2\left[{\rm d}r^2+\sinh^2r\left({\rm d}\theta^2+\sin^2\theta\,{\rm d}\varphi^2\right)\right].
\label{veinte}
\end{equation}
The invariant integration measure ${\rm d}\mu(r, \theta,\varphi)$ corresponding to the above metric is
\begin{equation}
{\rm d}\mu(r, \theta,\varphi)=\frac{1}{R_0^3}\sqrt{g}\,{\rm d}^3x=\sinh^2r\,\sin\theta\,{\rm d}r\,{\rm d}\theta\,{\rm d}\varphi,
\label{veintiuno}
\end{equation}
while the corresponding Laplacian operator on $\mathbb{H}^3$ reads
$$
\nabla^2=\frac{1}{\sqrt{g}}\partial_{i}\left(\sqrt{g}g^{ij}\partial_j\right)
$$
\begin{equation}
=\frac{1}{R_0^2\sinh^2 r}\left[\frac{\partial}{\partial r}\left(\sinh^2r\frac{\partial}{\partial r}\right)+\frac{1}{\sin\theta}\frac{\partial}{\partial\theta}\left(\sin\theta\frac{\partial}{\partial\theta}\right)+\frac{1}{\sin^2\theta}\frac{\partial^2}{\partial\varphi^2}\right].
\label{veintidos}
\end{equation}
We can reexpress the above as
\begin{equation}
\nabla^2=\frac{1}{R_0^2}\left[\frac{1}{\sinh^2r}\frac{\partial}{\partial r}\left(\sinh^2r\,\frac{\partial}{\partial r}\right)-\frac{{\bf L}^2}{\sinh^2r}\right],
\label{tres}
\end{equation}
where ${\bf L}^2$ is the square of the usual angular momentum operator on the 2--dimensional sphere $\mathbb{S}^2$. This allows one to solve the Laplacian eigenvalue equation $\nabla^2\psi=\lambda\psi$ by separation of variables: $\psi(r,\theta, \varphi)=R_{\lambda l}(r)Y_{lm}(\theta, \varphi)$, with $Y_{lm}(\theta,\varphi)$ a spherical harmonic on $\mathbb{S}^2$. The Laplacian eigenvalue equation is thus reduced to the ordinary differential equation
\begin{equation}
\frac{1}{R_0^2\sinh^2r}\left[\frac{{\rm d}}{{\rm d}r}\left(\sinh^2r\,\frac{{\rm d}}{{\rm d}r}\right)-l(l+1)\right]R_{\lambda l}(r)=\lambda R_{\lambda l}(r),
\label{cinco}
\end{equation}
hereafter termed {\it the radial wave equation}\/. One finds the continuous spectrum \cite{LAX}
\begin{equation}
\lambda=-\frac{1}{R_0^2}(k^2+1),\qquad k\in[0,\infty).
\label{seis}
\end{equation}
We will henceforth denote the radial eigenfunctions $R_{\lambda l}(r)$ as $R_{k l}(r)$. 

Imposing regularity of the solution of Eq. (\ref{cinco}) at $r=0$ yields the 1--dimensional eigenspace generated by the eigenfunction \cite{LIMIC}
$$
R_{k l}(r)=N_{k l}^{-1/2}\,\left(\sinh r\right)^l\,\left(\cosh r\right)^{{\rm i}k-l-1}
$$
\begin{equation}
\times F\left(\frac{l+1-{\rm i}k}{2},\frac{l+2-{\rm i}k}{2};l+\frac{3}{2};\tanh^2 r\right).
\label{siete}
\end{equation}
Above, $F(a,b;c;z)$ is the Gauss hypergeometric function \cite{GRADSHTEYN}; the precise value of the normalisation constant $N_{k l}$ can be looked up in ref. \cite{LIMIC}. A second, linearly independent solution of Eq. (\ref{cinco}) is singular at $r=0$, hence discarded.

Summarising, the Laplacian eigenfunctions $\psi_{klm}(r,\theta, \varphi)=R_{k l}(r)\,Y_{lm}(\theta,\varphi)$ are orthonormal
\begin{equation}
\int_{\mathbb{H}^3}{\rm d}\mu(r,\theta, \varphi)\,\psi^*_{klm}(r,\theta, \varphi)\psi_{k'l'm'}(r,\theta, \varphi)=\delta(k-k')\delta_{ll'}\delta_{mm'}
\label{doce}
\end{equation}
and complete
$$
\int_0^{\infty}{\rm d}k\,\sum_{l=0}^{\infty}\sum_{m=-l}^l\psi^*_{klm}(r,\theta, \varphi)\psi_{klm}(r',\theta', \varphi')
$$
\begin{equation}
=\frac{1}{\sinh^2 r\sin\theta}\delta(r-r')\delta(\theta-\theta')\delta(\varphi-\varphi').
\label{trece}
\end{equation}

As already done in refs. \cite{NOSALTRES, JMI} we will restrict our attention to radially symmetric states, {\it i.e.}\/, to $l=0$, $m=0$. Let us denote the $\psi_{k00}(r)$ more simply by $\psi_k(r)$. Then the radially symmetric wavefunctions are given by
\begin{equation}
\psi_k(r)=\left(\cosh r\right)^{{\rm i}k-1} F\left(\frac{1-{\rm i}k}{2},\frac{2-{\rm i}k}{2}; \frac{3}{2};\tanh^2 r\right),
\qquad k\in[0,\infty),
\label{catorce}
\end{equation}
modulo a normalisation factor to be specified later.

\subsection{Radially symmetric eigenfunctions from ladder operators}

The radially symmetric eigenfunctions (\ref{catorce}), elegant though they are, turn out to be somewhat awkward to work with. Fortunately a more practical expression for the solutions (\ref{catorce}) can be given along the lines first presented in ref. \cite{INFELD}. In order to obtain it we follow the method already applied in ref. \cite{JMI} to the case of spherical space $\mathbb{S}^3$, duly adapted to hyperbolic space $\mathbb{H}^3$. 

By inspection, the function defined by 
\begin{equation}
f_q(r)=\frac{\sinh(qr)}{\sinh r}, \qquad q\in\mathbb{C}
\label{setenta}
\end{equation}
satisfies the radial equation (\ref{cinco}) with $l=0$ whenever $q=\pm {\rm i}k$. Hence we have a family of states with $l=0$ given by
\begin{equation}
f_{{\pm}{\rm i}k}(r)=\pm {\rm i}\,\frac{\sin(kr)}{\sinh r}, \qquad k\in[0, \infty).
\label{ochentatres}
\end{equation}
Dropping the irrelevant factor $\pm{\rm i}$ let us consider
\begin{equation}
R_{k0}(r)=\frac{\sin(kr)}{\sinh r}.
\label{cuatrocientoscinco}
\end{equation}
We claim that the function defined as\footnote{We deliberately use the same notation $R_{kl}$ as in Eq. (\ref{siete}) because, as will be shown below, the functions (\ref{ochentados}) are indeed the same family of functions as given in Eq. (\ref{siete}), modulo an irrelevant proportionality factor. }
\begin{equation}
R_{kl}(r)=\sinh^lr\left(\frac{1}{\sinh r}\frac{{\rm d}}{{\rm d}r}\right)^l  R_{k0}(r)
\label{ochentados}
\end{equation}
satisfies the radial equation (\ref{cinco}) with an arbitrary value of $l\in\mathbb{N}$.

In order to prove the above statement we return to the radial Eq. (\ref{cinco}), which we reexpress as
\begin{equation}
\frac{{\rm d}^2  R_{kl}}{{\rm d}r^2}+2\coth r\frac{{\rm d}  R_{kl}}{{\rm d}r}+\left[(k^2+1)-\frac{l(l+1)}{\sinh^2r}\right]  R_{kl}=0,
\label{cincobis}
\end{equation}
after using Eq. (\ref{seis}). The change of dependent variable
\begin{equation}
R_{kl}(r)=\sinh^lr\,  C_{kl}(r)
\label{setentatres}
\end{equation}
turns Eq. (\ref{cincobis}) into
\begin{equation}
\frac{{\rm d}^2  C_{kl}}{{\rm d}r^2}+2(l+1)\coth r\,\frac{{\rm d}  C_{kl}}{{\rm d}r}+\left[l(l+2)+(k^2+1)\right]  C_{kl}(r)=0.
\label{setentacuatro}
\end{equation}
Differentiating (\ref{setentacuatro}) once more one arrives at
$$
\frac{{\rm d}^3  C_{kl}}{{\rm d}r^3}+2(l+1)\coth r\,\frac{{\rm d}^2  C_{kl}}{{\rm d}r^2}
$$
\begin{equation}
+\left[l(l+2)+(k^2+1)-2(l+1)\,{\rm csch}^2 r\right]\frac{{\rm d}  C_{kl}}{{\rm d}r}=0.
\label{setentacinco}
\end{equation}
We claim that the ansatz 
\begin{equation}
\frac{{\rm d}  C_{kl}}{{\rm d}r}=\sinh r\,  C_{k,l+1}
\label{ochentacuatro}
\end{equation}
for the third--order equation (\ref{setentacinco}) yields a simple solution of the second--order equation (\ref{setentacuatro}).
Indeed, substitution of (\ref{ochentacuatro}) into (\ref{setentacinco}) produces
\begin{equation}
\frac{{\rm d}^2  C_{k,l+1}}{{\rm d}r^2}+2(l+2)\coth r\,\frac{{\rm d}  C_{k, l+1}}{{\rm d}r}+\left[(l+1)(l+3)+(k^2+1)\right]  C_{k,l+1}=0,
\label{ochentacinco}
\end{equation}
which coincides with the radial Eq. (\ref{setentacuatro}) after the replacement $l\to l+1$. Thus the operator
\begin{equation}
O=\frac{1}{\sinh r}\frac{{\rm d}}{{\rm d}r}
\label{ochentaseis}
\end{equation}
acts as a ladder operator, {\it i.e.}\/, 
\begin{equation}
O\left(  C_{kl}(r)\right)=  C_{k,l+1}(r).
\label{ochentaseisbis}
\end{equation}
Therefore Eq. (\ref{ochentados}) does provide the solution to the radial Eq. (\ref{cincobis}) for arbitrary values of $k\in[0,\infty)$ and $l\in\mathbb{N}$; it is the only regular solution at $r=0$.  A second, linearly independent solution of Eq. (\ref{cincobis}) does exist, but it is singular at $r=0$. 

Turning now our attention to the normalisation of the radial eigenstates (\ref{ochentados}), on general grounds the orthogonality property \begin{equation}
\langle R_{kl}\vert R_{k'l}\rangle={\cal N}_{kl}\delta(k-k'), \qquad k,k'\in[0,\infty)
\label{cuatrocientosiete}
\end{equation}
will hold for some value of ${\cal N}_{kl}\in\mathbb{C}$.  We do not need the normalisation factor ${\cal N}_{kl}$ in all generality, but only its value for the radially symmetric states $l=0$. A simple computation proves that the eigenfunctions
\begin{equation}
\psi_k(r)=\frac{1}{\sqrt{\pi}}\,R_{k0}(r)=\frac{1}{\sqrt{\pi}}\frac{\sin(kr)}{\sinh r}, \qquad k\in[0,\infty)
\label{cuatrocientosdoce}
\end{equation}
are orthonormal with respect to the scalar product on the left--hand side of Eq. (\ref{doce}):
\begin{equation}
\langle\psi_k\vert\psi_{k'}\rangle=\delta(k-k'), \qquad k,k'\in[0,\infty).
\label{cuatrocientostrece}
\end{equation}
The $\psi_k(r)$ above span a complete, orthonormal set within the subspace of radially symmetric Laplacian eigenfunctions.

We have deliberately used the same notation $\psi_k(r)$ for the eigenfunctions of Eqs. (\ref{cuatrocientosdoce}) and (\ref{catorce}), because they are indeed the same set of functions (modulo normalisation). The reason for this is that the subspace of solutions to the radial Eq. (\ref{cinco}) that are regular at $r=0$ is 1--dimensional. In fact the following remarkable identity holds true:
\begin{equation}
 F\left(\frac{1-{\rm i}k}{2},\frac{2-{\rm i}k}{2}; \frac{3}{2};\tanh^2 r\right)=\frac{\sin(kr)\left(\cosh r\right)^{1-{\rm i}k}}{k\sinh r} .
\label{cuatrocientoscatorce}
\end{equation}
Despite the vastly different appearance of the functions on the two sides, the above identity holds for all $r\in(0,\infty)$ and all $k\in[0,\infty)$ thanks to Eq. 9.121.19 of ref. \cite{GRADSHTEYN}, after an appropriate analytic continuation of the latter to the complex plane.

\section{Evaluation of the cosmological constant and the Boltzmann entropy}

\subsection{Operators}

The analysis of  ref. \cite{JMI} suggests considering the operator 
\begin{equation}
\Lambda(r)=\frac{{C}_{\Lambda}}{R_U^2}\frac{1}{\sinh^2r}
\label{sesenta}
\end{equation}
as representing the cosmological constant. Above, ${C}_{\Lambda}$ is a dimensionless numerical constant left undetermined by our arguments, while the radius of the Universe $R_U$ provides the necessary dimensions because the radial coordinate $r$ is dimensionless. The same analysis of ref. \cite{JMI} suggests the entropy operator 
\begin{equation}
S(r)=\frac{{C}_{S}k_BMH_0 R_U^2}{\hbar}\sinh^2r. 
\label{sesentauno}
\end{equation}
As in Eq. (\ref{sesenta}), a  dimensionless numerical factor ${C}_{S}$ is left undetermined; however we will see that reasonable results are obtained upon setting both ${C}_{\Lambda}$ and ${C}_S$ to be of order unity. The physical constants appearing above are Boltzmann's constant $k_B$; the total mass $M$ (baryonic and dark) of the observable Universe; the radius $R_U$ of the observable Universe; Hubble's constant $H_0$. We draw their experimentally determined values from ref. \cite{PLANCK}.

\subsection{Action on the subspace of radially symmetric states}

Next we evaluate the matrix representing the cosmological--constant operator (\ref{sesenta}) within the subspace of radially symmetric states (\ref{cuatrocientosdoce}). We find
\begin{equation}
\langle \psi_{k}\vert \Lambda(r)\vert \psi_{k'}\rangle=4\frac{{C}_{\Lambda}}{R_U^2}\int_0^{\infty}{\rm d}r\,\frac{\sin(kr)\sin(k'r)}{\sinh^2r}, \qquad k,k'\in[0,\infty).
\label{cuatrocientosonce}
\end{equation}
Although the integral can be evaluated explicitly, the result is not very inspiring. In order to get a feeling for the orders of magnitude involved let us consider the diagonal elements, {\it i.e.}\/, let us take $k'=k$:
\begin{equation}
\langle\psi_k\vert\Lambda(r)\vert\psi_k\rangle=\frac{{C}_{\Lambda}}{R_U^2}\left[-1+2k\pi\coth(k\pi)\right], \qquad k\in[0,\infty).
\label{cuatrocientosveinte}
\end{equation}
The above is a monotonically increasing function of $k\in[0,\infty)$. Inserting the radius $R_U=4.4\times 10^{26}$ m and setting $C_{\Lambda}=1$ for simplicity, the best fit to the experimentally measured value of the cosmological constant $\Lambda=1.1\times 10^{-52}{\rm m}^{-2}$ is attained around $k=4$.

We turn now to the matrix representing the entropy operator (\ref{sesentauno}). It reads
\begin{equation}
\langle\psi_k\vert S(r)\vert\psi_{k'}\rangle=4\frac{C_Sk_BMH_0R_U^2}{\hbar}\int_0^{\infty}{\rm d}r\,\sinh^2r\sin(kr)\sin(k'r), \quad k,k'\in[0,\infty).
\label{cuatrocientosveintedos}
\end{equation}
As might have been expected, the above integral is infrared divergent because hyperbolic space is noncompact.  We can however regularise it by the introduction of $R_U$ as an infrared cutoff; this we can implement as follows. In the metric (\ref{veinte}) the radial coordinate $r$ is dimensionless, the dimensions of length arising from the factor $R_0$. Setting $R_0=R_U$, the coordinate value $r=1$ corresponds to the physical distance $R_U$. Within the radially symmetric subspace of Hilbert space, the regularised entropy operator is represented by the matrix
\begin{equation}
\langle\psi_k\vert S(r)\vert\psi_{k'}\rangle_{\rm reg}=4\frac{C_Sk_BMH_0R_U^2}{\hbar}\int_0^{1}{\rm d}r\,\sinh^2r\sin(kr)\sin(k'r), \; k,k'\in[0,\infty).
\label{cuatrocientosveintetres}
\end{equation}
Again restricting ourselves to the diagonal elements we find
\begin{equation}
\langle\psi_k\vert S(r)\vert\psi_{k}\rangle_{\rm reg}
=\frac{C_Sk_BMH_0R_U^2}{2\hbar}\,f(k),
\label{cuatrocientosveinticuatro}
\end{equation}
where the function $f(k)$ is defined by
\begin{equation}
f(k)=-\frac{1}{k^2+1}\left[\sinh(2)\cos(2k)+k\cosh(2)\sin(2k)\right]+\frac{\sin(2k)}{k}-2+\sinh(2).
\label{quinientos}
\end{equation}
The function $f(k)$ oscillates around the value $-2+\sinh(2)$, performing an infinite number of damped oscillations with a monotonically decreasing amplitude. The maximum amplitude is reached in the first oscillation (when $k\in(0,2)$), with $f_{\rm max}\simeq 3$. 

In order to verify that our model complies with the holographic principle, we replace $f(k)$ by its maximum value, $f_{\rm max}\simeq 3$, thus obtaining the estimate
\begin{equation}
\langle\psi_k\vert S(r)\vert\psi_{k}\rangle_{\rm reg}\leq\frac{C_Sk_BMH_0R_U^2}{2\hbar}3.
\label{cuatrocientosveinticinco}
\end{equation}
{}Further picking $C_S=2/3$ and substituting the known values of the cosmological data \cite{PLANCK}, we conclude that our results are indeed
compatible with the upper bound $10^{123}k_B$ set by the holographic principle:
\begin{equation}
\langle\psi_k\vert S(r)\vert\psi_{k}\rangle_{\rm reg}\leq 10^{123}k_B.
\label{quinientosuno}
\end{equation}

\section{Discussion}\label{conclusiones}

We have computed the cosmological constant $\Lambda$ and the Boltzmann entropy $S$ of a hyperbolic Universe $\mathbb{R}\times\mathbb{H}^3$ in the Newtonian limit. To this end we have modelled the cosmological fluid as the quantum probability fluid (a perfect fluid) of a free, nonrelativistic particle governed by Schroedinger quantum mechanics \cite{CABRERA, NOSALTRESTRIS}. This particle is assumed to possess a mass $M$ equal to that of the total mass (baryonic plus dark) of the observable Universe. The particle being free, its energy eigenstates are eigenfunctions of the Laplacian operator $\nabla^2$ on the spatial manifold $\mathbb{H}^3$.  A complete, orthonormal set of such eigenfunctions $\psi_{klm}$ (Eqs. (\ref{doce}), (\ref{trece})) can been borrowed from the existing literature \cite{LIMIC}; see also ref. \cite{MOSCHELLA}. However, upon restriction to radially symmetric states for isotropy, we have found it more convenient to work with the eigenfunctions $\psi_k$ of Eq. (\ref{cuatrocientosdoce}), used thereafter. 

Next, quantum operators $\Lambda$ and $S$ (respectively given by Eqs. (\ref{sesenta}) and (\ref{sesentauno}))  can be identified, such that they  represent the cosmological constant and the Boltzmann entropy of this Universe. The specific choice made for the entropy operator $S$ is motivated by the entropic approach to gravity of ref. \cite{VERLINDE1}.  In the Newtonian limit of this latter scenario, gravitational equipotential surfaces qualify as isoentropic surfaces.  

We have finally evaluated the expectation values of the two quantum operators $\Lambda$ and $S$ in the radially symmetric quantum states $\psi_k$ representing the matter contents of our hyperbolic Universe; our results are presented in Eqs. (\ref{cuatrocientosveinte}) and (\ref{cuatrocientosveinticuatro}). Specifically, the expectation value $\langle\psi_k\vert S\vert\psi_k\rangle$ measures the gravitational entropy of the Universe when the matter it contains finds itself in the quantum state $\psi_k$, while $\langle\psi_k\vert\Lambda\vert\psi_k\rangle$ measures the corresponding cosmological constant.

The radial quantum number $k\in[0,\infty)$ is (proportional to) the modulus of the momentum. We find that the best fit to the experimentally determined value of $\Lambda$ occurs around $k=4$. It is also worth remarking that the corresponding $\langle\psi_k\vert S\vert\psi_k\rangle$ saturates, but does not violate, the upper bound set by the holographic principle on the maximal entropy content of the Universe. These conclusions are in line with the corresponding ones drawn in flat space \cite{NOSALTRES} and in spherical space \cite{JMI}.

We recall that both ultraviolet ($r\to 0$) and infrared ($r\to\infty$) regularisations were needed in flat space \cite{NOSALTRES} in order to make physical sense of the expectation values $\langle\psi_k\vert\Lambda\vert\psi_k\rangle$ and $\langle\psi_k\vert S\vert\psi_k\rangle$. On the contrary, spherical space \cite{JMI} required no regularisation at all, neither infrared nor ultraviolet. Finally hyperbolic space as worked out here only required infrared regularisation. Thus nonvanishing curvature does away with the need for ultraviolet regularisation, but infrared regularisation remains necessary whenever space is noncompact.  

To round up our discussion it is instructive to discuss the eigenvalues $\lambda_k$ of the Laplacian $\nabla^2$ as functions of the radial quantum number $k$, and also in their dependence on the scalar curvature of the spatial manifold considered. For the cases when the latter is flat space $\mathbb{R}^3$ \cite{NOSALTRES}, spherical space $\mathbb{S}^3$ \cite{JMI}, and hyperbolic space $\mathbb{H}^3$ as analysed here, the following spectra are found in the literature (normalising the radii $R_0$ of the manifolds $\mathbb{S}^3$ and $\mathbb{H}^3$ to unity):
\begin{equation}
\begin{tabular}{| c | c | c |}\hline
$$ & eigenvalues of $\nabla^2$    &  range of $k$\\ \hline
$\mathbb{R}^3$  & $-k^2$  & $k\in[0,\infty)$\\ \hline
$\mathbb{S}^3$  & $-k(k+2)$  & $k\in\mathbb{N}$\\ \hline
$\mathbb{H}^3$  & $-(k^2+1)$  & $k\in[0,\infty)$\\ \hline
\end{tabular}
\label{trumptheloser}
\end{equation}

As was to be expected, only the compact manifold $\mathbb{S}^3$ leads to a discrete spectrum, while the two noncompact manifolds $\mathbb{R}^3$ and $\mathbb{H}^3$ possess continuous spectra. In all three cases, $-\nabla^2$ is a nonnegative operator as befits a kinetic energy. One can regard terms in $\lambda_k$ additional to $-k^2$ as due to nonvanishing scalar curvature. Thus curvature corrections to the flat case cause the kinetic energy to increase with respect to their flat--space counterpart. In flat space and also in spherical space, the state with zero momentum $k=0$ also has zero kinetic energy. However, hyperbolic space allows a nonvanishing kinetic energy even when the momentum $k$ vanishes. Thus the hyperbolic Laplacian is not just nonnegative, but strictly positive; this is also a curvature effect.

{\bf Acknowledgements} This research was supported by grant no. RTI2018-102256-B-I00 (Spain).

\end{document}